\documentclass[prb,superscriptaddress,reprint,showpacs,amssymb,amsmath,floatfix,aps]{revtex4-1}

\usepackage{graphicx}

\usepackage{color}          

\begin{document}

\title{Imaging and Manipulation of Skyrmion Lattice Domains in Cu$_2$OSeO$_3$}

\author{S. L. Zhang}
\affiliation{Clarendon Laboratory, Department of Physics, University of Oxford, Parks Road, Oxford, OX1~3PU, United Kingdom}

\author{A. Bauer}
\affiliation{Physik Department, Technische Universit\"at M\"unchen, James-Franck-Strasse 1, 85748 Garching, Germany}

\author{H. Berger}
\affiliation{Crystal Growth Facility, Ecole Polytechnique F\'{e}d\'{e}rale de Lausanne (EPFL), CH-1015 Lausanne, Switzerland}

\author{C. Pfleiderer}
\affiliation{Physik Department, Technische Universit\"at M\"unchen, James-Franck-Strasse 1, 85748 Garching, Germany}

\author{G. \surname{van~der~Laan}}
\affiliation{Magnetic Spectroscopy Group, Diamond Light Source, Didcot, OX11~0DE, United Kingdom}

\author{T. Hesjedal}
\affiliation{Clarendon Laboratory, Department of Physics, University of Oxford, Parks Road, Oxford, OX1~3PU, United Kingdom}
\email{Thorsten.Hesjedal@physics.ox.ac.uk}

\date{\today}

\begin{abstract}
Nanoscale chiral skyrmions in noncentrosymmetric helimagnets are promising binary state variables in high-density, low-energy nonvolatile memory. Nevertheless, they normally appear in an ordered, single-domain lattice phase, which makes it difficult to write information unless they are spatially broken up into smaller units, each representing a bit. Thus, the formation and manipulation of skyrmion lattice domains is a prerequisite for memory applications. Here, using an imaging technique based on resonant magnetic x-ray diffraction, we demonstrate the mapping and manipulation of skyrmion lattice domains in Cu$_2$OSeO$_3$. The material is particularly interesting for applications owing to its insulating nature, allowing for electric field-driven domain manipulation.
\end{abstract}

\maketitle

Skyrmions are particle-like magnetization swirls in magnetic materials,\cite{Pf_MnSi_Science_09,Tokura_review_skyrmion_Natnano_13,Nanjing_Theory_artificial_skyrmion_PRL_13,Jiang_blowing_bubble_Science_15,CNRS_[CoFeB/Pt]n_STXM_Natphys_15,White_GaV4S8_Natmater_15,CNRS_Ta/Pt/Co/MgO/Ta_PEEM_Natnano_16} which are promising candidates for advanced magnetic memory applications.\cite{Tokura_review_skyrmion_Natnano_13}
In these memory schemes, individual skyrmions are used to encode the binary information `1' and `0', e.g., via their presence\cite{Hamburg_Ir/FePd_write_Science_13,VA,MIT_[Ta/Pt/Co]n_STXM_Natmater_16} or their internal structures.\cite{Cu2OSeO3_multidomain_Nanoletters_16,Diaz_helicity_theory_IOP_16} Skyrmions are topological objects, which makes them robust against superparamagnetism.
Consequently, it should be possible to reduce the bit size beyond the limits of conventional ferromagnetic memory. Further, the energy required to manipulate skyrmions is several orders of magnitude less than domain wall-based ferromagnetic memory.\cite{Pf_MnSi_STT_rotation_Science,Rosch_MnSi_emergent_NatPhys,Tokura_CuOSeO-MnSi_ratchet_Natmater_14}

Prominent skyrmion-carrying materials are noncentrosymmetric helimagnets, such as MnSi,\cite{Pf_MnSi_Science_09} Fe$_{1-x}$Co$_x$Si,\cite{Pf_FeCoSi_SANS_PRB_10} FeGe,\cite{Tokura_FeGe_STT_NatComm_12} Cu$_2$OSeO$_3$,\cite{Tokura_CuOSeO_LTEM_Science_12,Tokura_CuOSeO_REXS_PRL_14} and $\beta$-type Co$_8$Zn$_8$Mn$_4$.\cite{CoMnZn} In such materials, the broken crystalline inversion symmetry induces the Dzyaloshinskii-Moriya interaction (DMI), leading to periodic, incommensurate, modulated spiral spin structures.
Assisted by thermal fluctuations at finite temperature and in an external field, the topologically protected skyrmion lattice phase forms, consisting of chiral skyrmions.\cite{Pf_MnSi_Science_09}
The advantages of chiral skyrmions for memory devices are twofold. First, compared to magnetic bubbles with similar topological spin configurations,\cite{Bubble_review_Scinece_80} skyrmions are more robust and can be manipulated with ease.
The size of an individual skyrmion in these materials is usually between 3-100~nm,\cite{Pf_MnSi_Science_09,2010:Yu:Nature,YuFeGe2011,2011:Kanazawa} i.e., smaller than other types of skyrmions.\cite{Bubble_review_Scinece_80,Tokura_BaFeScMgO_reversal_PNAS_12,Nanjing_[Co/Pt]n_MOKE_PRB_14,Jiang_blowing_bubble_Science_15,CNRS_[CoFeB/Pt]n_STXM_Natphys_15,Berkeley_[Fe/Ni]n_SPLEEM_APL_15,CNRS_Ta/Pt/Co/MgO/Ta_PEEM_Natnano_16,MIT_[Ta/Pt/Co]n_STXM_Natmater_16}
Second, the skyrmion phase is a rigid, hexagonally-ordered periodic lattice, resulting in an equal spacing between neighboring skyrmions across the entire sample.\cite{Pf_MnSi_long_range,Tokura_review_skyrmion_Natnano_13}
As a consequence, in a racetrack-like memory scheme,\cite{Parkin_racetrack_Science_08} no extra effort is needed to assure that they keep their distance, as the control of the skyrmion-skyrmion distance is otherwise experimentally rather challenging.\cite{Hamburg_Ir/FePd_write_Science_13,Jiang_blowing_bubble_Science_15,MIT_[Ta/Pt/Co]n_STXM_Natmater_16}

The skyrmion lattice phase in noncentrosymmetric helimagnets is usually a long-range-ordered state, with the correlation length reaching hundreds of micrometers.\cite{Pf_MnSi_long_range}
This largely limits the applicability of this type of skyrmion order for device applications. Therefore, the formation and manipulation of skyrmion lattice domains, which break the long-range order, is an indispensable step towards skyrmion-based racetrack memories.\cite{SkyRace_Tomasello}
We have recently reported the observation of a multidomain skyrmion lattice state on the surface of Cu$_2$OSeO$_3$ in a magnetic diffraction experiment, created by tilting the magnetic field away from the major crystalline axis.\cite{Cu2OSeO3_multidomain_Nanoletters_16,2016:Zhang-PRB} 
However, several key questions remained unanswered.
Most importantly, the size, shape, and distribution of the domains remained unknown, which are crucial pieces of information needed for designing skyrmion devices.

In this Letter, we use an imaging technique based on resonant elastic x-ray scattering (REXS) to map out the lateral distribution of skyrmion lattice domains. Moreover, we demonstrate that by tuning the tilt angle of the applied field, the size of the domains can be efficiently manipulated, which is an important step towards future applications.


\begin{figure*}[ht!]
	\begin{center}
		\includegraphics[width=16cm]{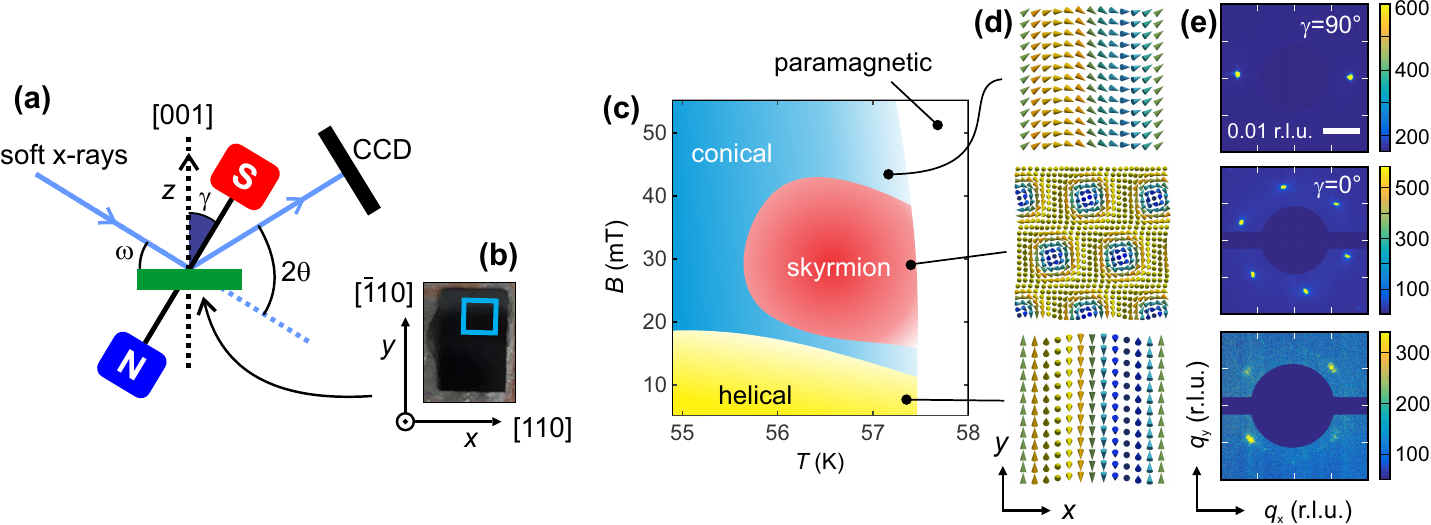}
		\caption{(a) Schematic of the REXS setup, showing the field tilt angle $\gamma$, defined with respect to the surface normal.
			(b) Photograph of the sample and its orientation. The blue square marks the real-space region ($1 \times 1$~mm$^2$) mapped using magnetic diffraction imaging, shown in Fig.\ \ref{fig:domains}.
			(c) Sketch of the magnetic phase diagram of Cu$_2$OSeO$_3$, as obtained by REXS.
			(d) Magnetization patterns and (e) corresponding REXS diffraction patterns in reciprocal space $(hk1)$ plane (experimental data) for the conical (50~K, 32~mT, $\gamma=90^\circ$), skyrmion (57~K, 32~mT, $\gamma=0^\circ$), and helical (15~K, 0~mT) phase, respectively. The sampling area is $500\times500~\mu\mathrm{m}^2$.}
		\vspace*{-0.5cm}
		\label{fig:setup}
	\end{center}
\end{figure*}

Figure \ref{fig:setup}(a) shows a schematic of the REXS setup used for the characterization of the magnetically ordered phases of Cu$_2$OSeO$_3$ as a function of applied magnetic field and temperature. A $\omega$-$2\theta$ geometry was used for the experiments, where $\omega$ is the angle of incidence of the x-rays and $2 \theta$ the scattering angle. The diffracted x-rays are detected with a charge-coupled device (CCD) camera in the ultrahigh vacuum scattering chamber RASOR on beamline I10 at the Diamond Light Source.\cite{RASOR_review} A magnetic field was applied to the sample whose strength and tilt angle, $\gamma$, with respect to the surface normal can be varied.
The incident, $\sigma$-polarized x-rays were at the resonance with the $L_3$ edge of Cu (931.25~eV), resulting in a wavelength of 13.3~\AA.
For Cu$_2$OSeO$_3$ with its relatively large lattice constant of 8.925~\AA, the (001) Bragg peak can be reached at $2 \theta \approx 96.5^\circ$.

In noncentrosymmetric $P2_13$ helimagnets, a `universal' magnetic phase diagram is observed that shows helical, conical, or ferrimagnetic order below the transition temperature $T_\mathrm{c}$ as a function of increasing field.\cite{Rosch_3D_Monte_Carlo_PRB_13}
Below $T_\mathrm{c}$ of $\sim$57.5~K for Cu$_2$OSeO$_3$, at finite fields, the skyrmion phase can be found. 
Figure \ref{fig:setup}(c) shows a schematic of the magnetic phase diagram for the Cu$_2$OSeO$_3$ bulk crystal as observed by REXS.\cite{Cu2OSeO3_multidomain_Nanoletters_16} 
The magnetization patterns corresponding to the helical, skyrmion, and conical phase are shown in Fig.\ \ref{fig:setup}(d), and the experimental REXS results in (e), respectively (from top to bottom).
The modulation wavevector for all three phases has a length of 0.0158-0.0162~reciprocal lattice units (r.l.u.), corresponding to a real-space modulation pitch of $\sim$56~nm, in agreement with the values reported in the literature.\cite{Pf_CuOSeO_PRL_12, Tokura_CuOSeO_LTEM_Science_12, Tokura_CuOSeO_rotation_PRB_12, White_Cu2OSeO3_E_rotation_IOP_12, Tokura_CuOSeO_REXS_PRL_14, White_Cu2OSeO3_LTEM_PNAS_15}
In the helical state, the weak cubic anisotropy locks the propagation wave vector along the $\langle$100$\rangle$ direction.
At 57~K and in an applied magnetic field of 32~mT, $\gamma=0^\circ$, the sharp six-fold-symmetric diffraction pattern emerges, which is a signature of the single-domain, long-range-ordered skyrmion lattice state.
One of the skyrmion wave vectors is along [010] due to the cubic anisotropy.\cite{Cu2OSeO3_multidomain_Nanoletters_16}

\begin{figure}[ht!]
	\begin{center}
		\includegraphics[width=8.5cm]{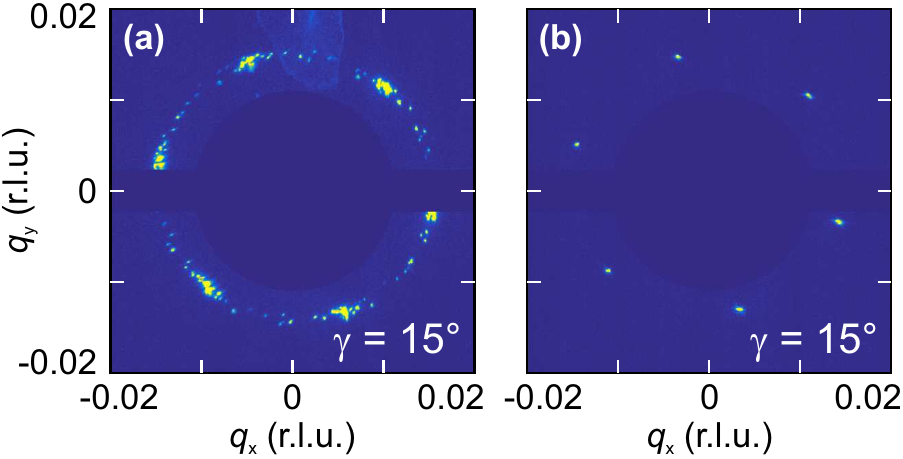}
		\caption{Reciprocal space map of the skyrmion lattice plane that is perpendicular to the external field, reached by field cooling down to 57~K in a field of 32~mT with $\gamma=15^\circ$. Area sampled by the beam: (a) $300\times300~\mu\text{m}^2$ and (b) $20~\mu\text{m}$ in diameter.
		The scattering intensity is in arbitrary units.}
		\vspace*{-1.0cm}
		\label{fig:focus}
	\end{center}
\end{figure}

By titling the magnetic field to $\gamma=17^\circ$, the skyrmion lattice breaks up into domains, resulting in a necklace-like diffraction pattern, as shown in Fig.\ \ref{fig:focus}(a).
The formation of domains is the result of the competing magnetic anisotropies.\cite{Cu2OSeO3_multidomain_Nanoletters_16}
Note that the diffraction pattern was obtained with an incident x-ray beam focused to an area of 300 $\times$ 300~$\mu$m$^2$ on the sample. Once the beam is skimmed down to an area of 20~$\mu$m in diameter (using a pinhole), the single domain state with its six-fold-symmetric pattern is recovered [cf.\ Fig.\ \ref{fig:focus}(b)]. 
This means that the domains are $>$$100\pi~\mu$m$^2$ in area, and that this beam spot can be used to map out the real-space domain pattern.

\begin{figure*}[ht!]
	\begin{center}
		\includegraphics[width=16cm]{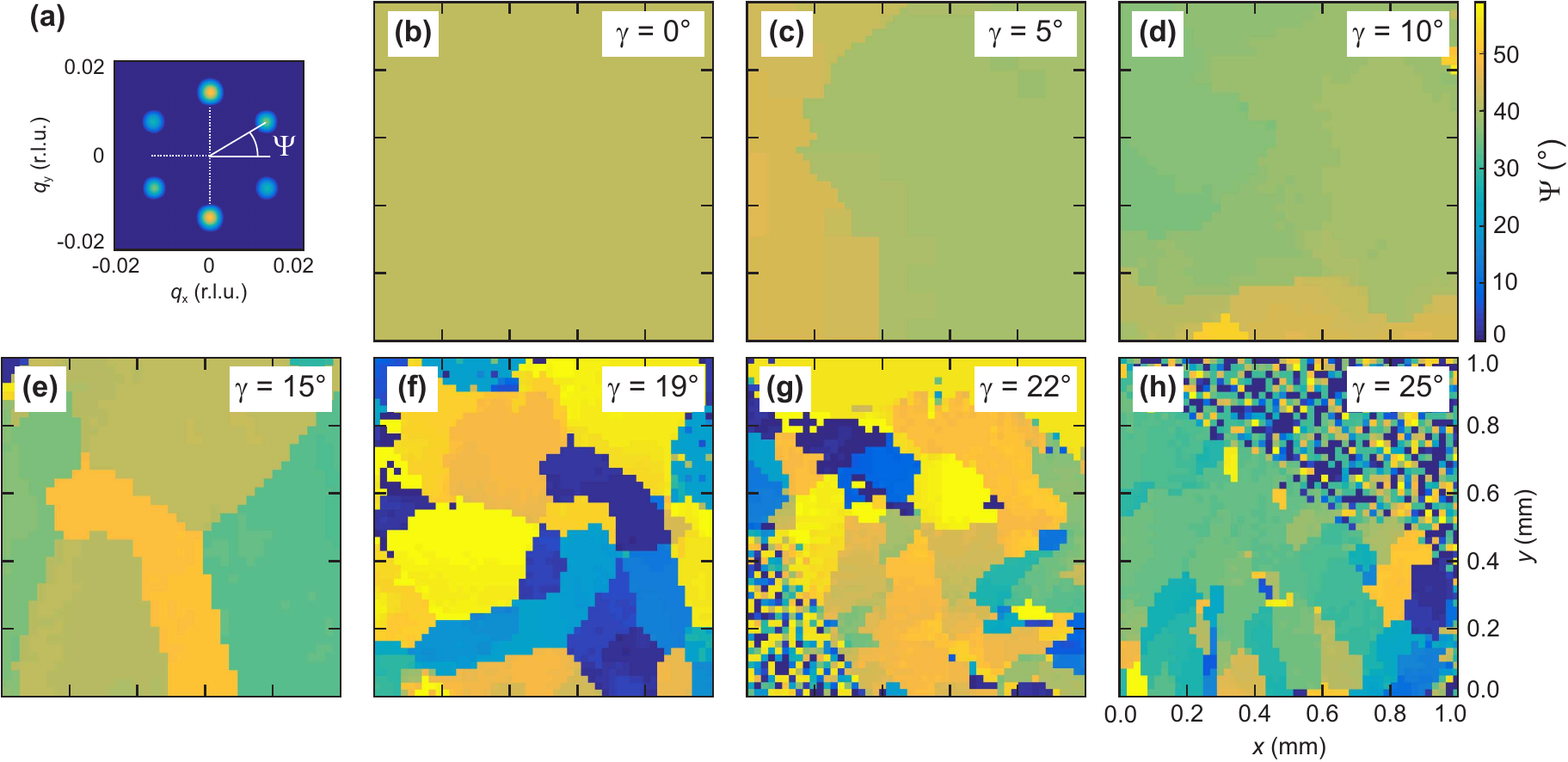}
		\caption{Scanning REXS images. (a) Definition of the lattice rotation angle $\Psi$ in the skyrmion diffraction pattern ($\Psi \in [0^\circ,60^\circ]$ as indicated by the color bar). The case of $\Psi=30^\circ$ is shown.
			(b-h) 1~mm $\times$ 1~mm raster scans showing in-plane rotational domain maps for the indicated field angle $\gamma$.}
		\vspace*{-0.9cm}
		\label{fig:domains}
	\end{center}
\end{figure*}

The sample shape and geometry is shown Fig.\ \ref{fig:setup}(b), in which the raster-scanned area ($1\times1$~mm$^2$) is indicated by the blue square.
The skyrmion lattice state is reached by field-cooling from 65~K (paramagnetic phase) down to 57~K in a field of 32~mT with the $\gamma$ angle as indicated in Fig.\ \ref{fig:domains}(b-h), followed by a 15~min waiting period. 
Each area scan image is composed of $50\times50$ pixels. 
For most of the pixels, a six-fold-symmetric, single domain diffraction pattern is observed. Its rotational state is characterized by the in-plane rotation angle $\Psi$, as shown Fig.\ \ref{fig:domains}(a). An angle of $\gamma =0^\circ$ corresponds to two of the six diffraction spots being aligned along the $q_x$-axis.
For some pixels, especially for larger field tilt angles $\gamma$, a multidomain state is found at the domain boundaries.
The evolution of the domain pattern as a function of field tilt angle is shown in Fig.\ \ref{fig:domains}(b-h).
For $\gamma=0^\circ$, a perfect long-range-ordered single skyrmion lattice domain is observed [cf.\ Fig.\ \ref{fig:domains}(b)], locked in one direction determined by the anisotropies.
When the field is slightly tilted ($\gamma = 5^\circ$), two domains start to emerge, which do not differ very much in $\Psi$.
As the field is further tilted, the domains become randomly oriented and the average size decreases (most strongly for $\gamma = 15^\circ \rightarrow 19^\circ$).
For $\gamma = 22^\circ$, a mosaic pattern is observed at the bottom-left corner of Fig.\ \ref{fig:domains}(g). This results from a lack of resolution, as determined by the x-ray beam size.
As the domain size decreases to less than the beam size, a multidomain diffraction pattern is observed across each pixel.
Note that the pinhole diameter can not be reduced much below 20~$\mu$m in our experimental configuration, as this will result in Fresnel diffraction (near-field condition giving cylindrical wave fronts), i.e., the plane wave approximation can no longer be applied for modeling the scattering process.

The shape of the domains is generally rather irregular, and their distribution random, suggesting that the domain formation is spontaneous and not governed by defect-pinning.
The domain pattern obtained for each $\gamma$ is stable over time, as confirmed by multiple scans of the same area.
Note, however, that for the same $\gamma$, if the temperature is increased above $T_\mathrm{c}$ and then lowered down to the skyrmion phase again, the domain image has no resemblance to the previous pattern.
The orientation of neighboring domains is similar and the system generally prefers the orientation of neighboring domains to differ by less than $10^\circ$. At higher tilt angles ($\gamma \geq 15^\circ$), the domain boundaries are sharp, however, for smaller angles the transition between domains becomes almost continuous, as can be seen in Fig.\ \ref{fig:domains}(d).

This `polycrystalline' appearance of a skyrmion lattice reflects the delicate balance of the system's competing interactions.
The isotropic exchange interaction, the anisotropic exchange (DMI) term, and the Zeeman energy compete
about the principal magnetic order, while anisotropy, demagnetization, and possibly the ferroelectric effect in the multiferroic material Cu$_2$OSeO$_3$ serve as perturbations which are responsible for the fine adjustment of the system's structure.
Skyrmions behave like quasi-particles and prefer to keep their close-packed order, i.e., less ferromagnetic-like space in-between them is preferred. However, if defects are introduced into the system, some region would have to become ferromagnetic, unless the skyrmions adapt their shape and size. 
As a result, the skyrmions in the defect zone can become elliptically distorted, as observed using Lorentz transmission electron microscopy (LTEM) on thinned-down FeGe$_{1-x}$Si$_x$ bulk samples.\cite{Tokyo_FeGeSi_LTEM_Scienceadv_16} 
Moreover, the close-packed ordering of the skyrmion lattice can have defects at the domain boundaries, e.g., `5-7 defects' where five or seven skyrmions are surrounding a reference skyrmion, instead of six.\cite{White_Cu2OSeO3_LTEM_PNAS_15} In this case, a line defect can appear, forming the domain boundary. Depending on its configuration, the neighboring domains can change their orientations and take an arbitrary angle, depending on the width of the defect-containing boundary. The thicker this boundary, the larger the relative rotation. If we relate this to our domain observations, it points towards the fact that for small $\gamma$, the domain boundary contains thin defect lines, leading to small relative rotations. With increasing $\gamma$, the number of defects increases across the domain boundary, giving rise to sharp rotational transitions.
Note that LTEM imaging, despite being immensely useful for studying the local defects of the skyrmions, is not able to map out the large-scale skyrmion structures like skyrmion lattice domains. Moreover, for systems with such a delicate energy balance as Cu$_2$OSeO$_3$, LTEM sample preparation-induced defects may affect the intrinsic surface domain structure (essentially a property of the bulk crystal). 
Therefore, the type and density of defects, and their influence on the observed domains, can be significantly different between bulk and thinned-down LTEM samples.

In summary, we have presented a study of the domain structure in the skyrmion lattice phase of Cu$_2$OSeO$_3$ using REXS imaging.
This method provides important information on the domain distribution, shape, size, and formation.
By tuning the field tilt angle, in-plane rotational domains in the hexagonally ordered skyrmion lattice phase can be generated. The lateral dimensions of the domains are for small tilt angles $>$20~$\mu$m. The domain transition is abrupt, expect for very small tilt angles, where the change is almost gradual.
The study of skyrmion domains may enable potential device applications making use of the skyrmion lattice state. 
In skyrmion lattice-based memory, the single-domain state must be broken up into domains, which each encode information in a spatially separated manner.
This was demonstrated using the field tilt angle as an additional handle.


The REXS experiments were carried out on beamline I10 at the Diamond Light Source, UK, under proposals SI-11784 and SI-12958. 
We thank the EPSRC for support under grant EP/N032128/1.
S.L.Z.\ and T.H.\ acknowledge financial support by the Semiconductor Research Corporation. A.B.\ and C.P.\ acknowledge financial support through DFG TRR80 and ERC AdG (291079, TOPFIT).



%

\end{document}